\documentclass[11pt,twoside]{article}
\usepackage{asp2010}

\resetcounters

\begin{document}

\title{Monte Carlo simulations of the luminosity function of hot white dwarfs}
\author{S. Torres,$^{1,2}$ 
        E. Garc\'{\i}a--Berro,$^{1,2}$, 
        J. Krzesinski,$^{3}$ and 
        S. J. Kleinman,$^4$}
\affil{$^1$Departament de F\'\i sica Aplicada, 
           Universitat Polit\`ecnica de Catalunya,  
           c/Esteve Terrades, 5,  
           08860 Castelldefels,  
           Spain}
\affil{$^2$Institute for Space Studies of Catalonia,
           c/Gran Capit\`a 2--4, Edif. Nexus 104,   
           08034  Barcelona, 
	   Spain}
\affil{$^3$Mt. Suhora Observatory, 
           Cracow Pedagogical University, 
           ul. Podchorazych 2, 
           30-084 Cracow, 
           Poland}
\affil{$^4$Gemini Observatory, 
           670 N. A'Ohoku Place, 
           Hilo, 
           HI 96720, 
           USA}

\begin{abstract}
We present a detailed Monte  Carlo simulation of the population of the
hot branch  of the white dwarf  luminosity function. We  used the most
up-to-date  stellar  evolutionary models  and  we  implemented a  full
description  of the observational  selection biases.   Our theoretical
results are compared with the  luminosity function of hot white dwarfs
obtained from  the Sloan  Digital Sky Survey  (SDSS), for both  DA and
non-DA white  dwarfs.  For  non-DA white dwarfs  we find  an excellent
agreement with the  observational data, while for DA  white dwarfs our
simulations  show some  discrepancies  with the  observations for  the
brightest   luminosity  bins,  those   corresponding  to   $L\ga  10\,
L_{\sun}$.
\end{abstract}

\section{Introduction}

Among other applications  --- see, for instance, the  recent review of
\citet{Althausetal10} --- the luminosity function and space density of
white  dwarfs  provide  interesting  constraints  on  the  local  star
formation  rate  and  history  of  the  Galactic  disk  in  the  Solar
neighborhood. However, to address  these questions a large white dwarf
sample of  known completeness  is required. The  SDSS has  provided us
with  such a sample,  and from  it a  reliable white  dwarf luminosity
function   for   hot  white   dwarfs   has   been  recently   obtained
\citep{Krz2009}.  This   white  dwarf  luminosity   function  has  the
interesting  particularity  that it  has  been  obtained using  solely
spectroscopically-confirmed white  dwarfs from the SDSS  DR4, and thus
constitutes an  excellent testbed  to check not  only the  white dwarf
cooling sequences at high luminosities,  but also our ability to model
reliably  some Galactic  inputs  necessary to  compute the  luminosity
function. Here we describe the results of a comprehensive set of Monte
Carlo  simulations aimed  to model  the hot  part of  the  white dwarf
luminosity function for both DA and non-DA stars.

\begin{figure}[ht]
\plotfiddle{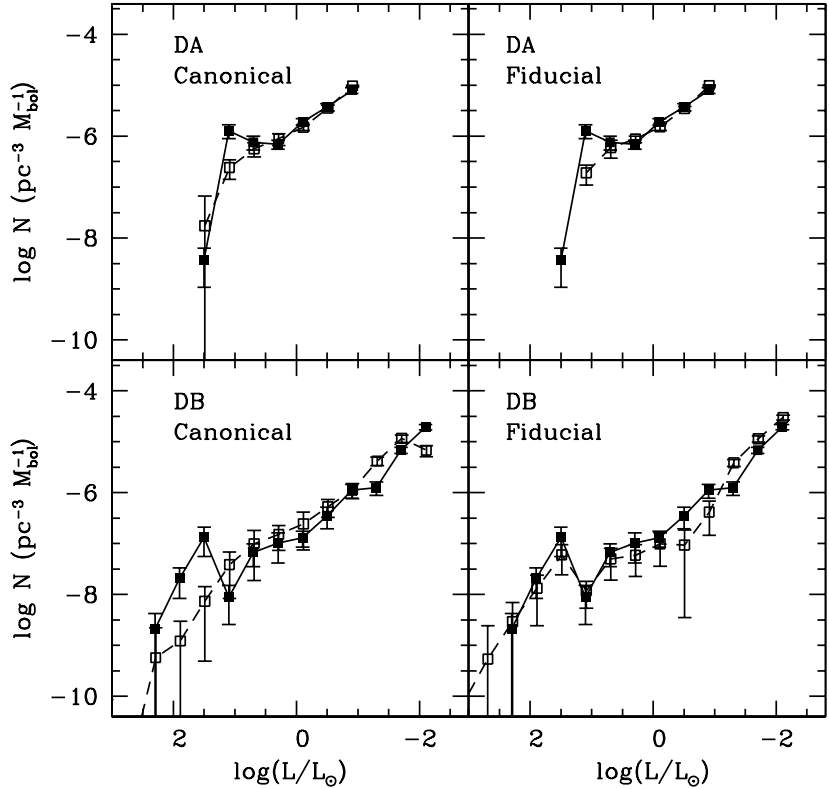}{10.0cm}{0}{64}{64}{-188}{-137}
\caption{White  dwarf luminosity  functions for  hot DA  and  DB white
  dwarfs for  two different models.  The solid line is  the luminosity
  function of \citet{Krz2009}, while  the dashed line shows the result
  of our simulations.}
\end{figure}

\section{The Monte Carlo simulator}

We simulated a synthetic population  of disk white dwarfs in the Solar
neighborhood in  a sphere of  3~kpc.  An extensive description  of our
Monte   Carlo   simulator   can    be   found   in   previous   papers
\citep{MC1,MC2,MC3}.  Thus, here we  only summarize the most important
inputs.  We adopted a disk  age of 10.5~Gyr, a constant star formation
rate      and      a      standard     initial      mass      function
\citep{Kroupa_2001}. Velocities were  obtained taking into account the
differential rotation of the Galaxy,  the peculiar velocity of the Sun
and a  dispersion law which depends  on the Galactic  scale height.  A
double exponential  profile was used, with a  scale heigth $h=250\,$pc
and a scale length $L=1.3\,$kpc.  Also, the initial-final relationship
of \citet{cataetal2008}  was adopted.  The  cooling sequences employed
depend  on the  mass of  the white  dwarf, $M_{\rm  WD}$.   If $M_{\rm
WD}\leq  1.1\,M_{\sun}$  a CO  core  was  adopted,  while  if  $M_{\rm
WD}>1.1\, M_{\sun}$  an ONe core  was used.  In  the case of  CO white
dwarfs with H-rich envelopes  we used the evolutionary calculations of
\citet{Renedo_10}, while for white dwarfs with ONe cores we used those
of  \citet{Althaus2007}.  For  H-deficient  white dwarfs  we used  the
cooling sequences  of \citet{Benvenuto1997}, which  correspond to pure
He  atmospheres, and the  bolometric corrections  of \citet{Bergeron}.
Finally,  we used  the same  selection criteria  employed to  cull the
observational  sample of  \citet{Krz2009}.   Specifically, only  white
dwarfs with  $g>14$ and a  fully de-reddened magnitude  $g_{\rm o}<19$
were selected, whereas the color cuts were $-1.5<(u-g)_{\rm o}< 0$ and
$-1.5<(g-r)_{\rm o}<0$.

\begin{figure}[t]
\plotfiddle{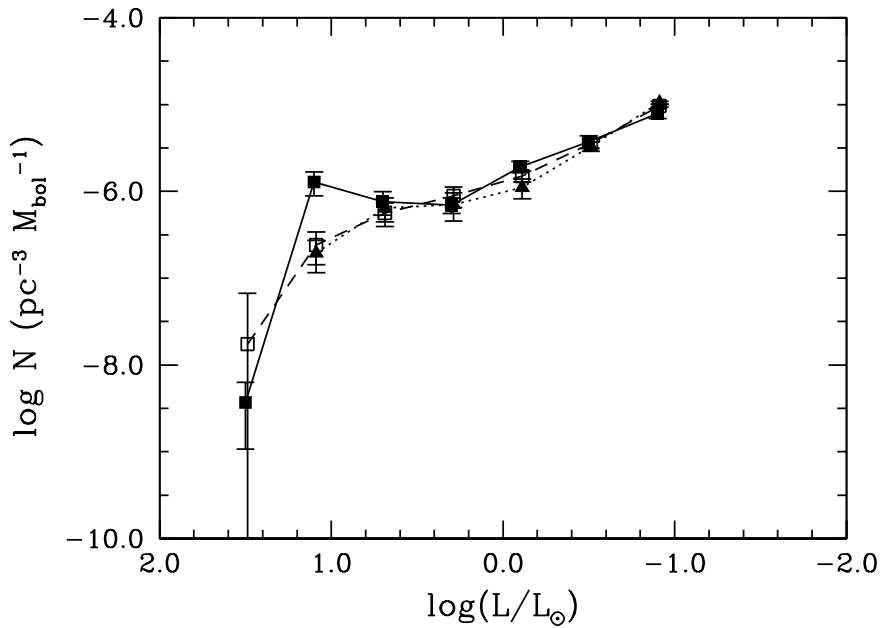}{9.0cm}{0}{66}{66}{-188}{-154}
\caption{White dwarf luminosity functions  for hot DA white dwarfs and
  different assumptions. The solid  line is the luminosity function of
  \citet{Krz2009}, the dashed line shows our fiducial model, while the
  dotted line represents the luminosity function when NLTE corrections
  and metal contamination is considered.}
\end{figure}

\section{Results}

In Fig.~1 we show several model luminosity functions for hot DA and DB
white dwarfs for two different  assumptions. In the canonical model we
adopt  a  DA/DB  ratio  $f_{\rm  DA/DB}=0.80$,  independently  of  the
effective temperature.  In a second simulation $f_{\rm DA/DB}$ depends
on  $T_{\rm eff}$,  as obtained  from the  SDSS  \citep{Krz2009}.  The
solid  lines  represent the  luminosity  function of  \citet{Krz2009},
while the  dashed lines show our simulated  luminosity functions.  The
upper panels  of Fig.~1 show that  the luminosity function  of hot DAs
barely depends on the DA/DB  ratio, whilst that of non-DA white dwarfs
depends sensitively on it.  Also,  it is clear that the agreement with
the observational data  is excellent for the model  in which the DA/DB
ratio  depends on $T_{\rm  eff}$.  Thus,  we adopt  this model  as our
fiducial one.

One puzzling  characteristic of the  observational luminosity function
of  hot white dwarfs  is the  existence of  a plateau  at luminosities
around $\log  (L/L_{\sun})\approx 1$.  In  a first attempt  to explore
its  origin,  we analyzed  the  effects  of  NLTE corrections  and  of
metallicity.  We adopted the  NLTE corrections of \citet{Napi1999} for
temperatures   $30,000\,{\rm  K}<T_{\rm  eff}<100,000\,{\rm   K}$  and
gravities $6.50 <\log g<9.75$.  Additionally we assumed that 50$\%$ of
hot DA  white dwarfs have  metals in their atmospheres.   This implies
that  their  effective  temperature  is  overestimated  by  20$\%$  at
$80\,000\,$K  and 0$\%$  at  $40,000\,$K.  The  results  are shown  in
Fig.~2.  The  dashed line  corresponds to our  fiducial model  and the
dotted  line to the  model in  which both  NLTE corrections  and metal
contamination have  been included.  Clearly,  these additional effects
have a small impact on the computed luminosity function.

Finally, to  assess if the plateau  of the luminosity  function of hot
DAs   could  correspond   to  a   recent  burst   of   star  formation
\citep{NohScalo1990},  we computed  several luminosity  functions with
different burst strengths occurring at various ages. However, we found
that  for   the  approximate   age  of  a   typical  white   dwarf  of
$0.6~M_{\sun}$ at $\log (L/L_{\sun})\approx  1$, the burst should have
occurred very recently. Thus, there are only two possible alternatives
left. Either the theoretical  cooling models are not entirely reliable
at  these   luminosities  ---  possibly  influenced   by  the  initial
conditions ---  or, instead,  since at this  luminosity the  number of
objects is  relatively small the observational error  bars inherent to
the    $(1/V_{\rm    max})$    method   have    been    underestimated
\citep{Geijoetal06},  and  the plateau  corresponds  to a  statistical
fluctuation.

\section{Conclusions}

We presented  a set of simulations  of the luminosity  function of hot
white  dwarfs.   Our  results  are  in excellent  agreement  with  the
observations  for  DB  stars,  while  the plateau  of  the  luminosity
function of DAs at  $\log (L/L_{\sun})\approx 1$ cannot be reproduced.
At  these luminosities  NLTE effects  and metal  contamination  play a
minor role in shaping the  luminosity function, and the plateau cannot
be explained by a burst of star formation.  Thus, it could be due to a
failure of the cooling models or to a statistical fluctuation.

\acknowledgements 
This research  was supported by AGAUR, by  MCINN grants AYA2011--23102
and AYA08-1839/ESP, by the European  Union FEDER funds, and by the ESF
EUROGENESIS project (grant EUI2009-04167).

\bibliography{torres}

\end{document}